\documentclass[aps,prl,twocolumn,superscriptaddress,showpacs,fleqn]{revtex4-1}

\usepackage{graphicx}
\usepackage{amsmath, amsfonts, amssymb, amsbsy}
\usepackage{MnSymbol}
\usepackage[usenames,dvipsnames]{xcolor}
\usepackage{marvosym}
\usepackage{sidecap}
\usepackage{ifsym}
\usepackage{color}

\newcommand{\beq} {\begin{equation}}
\newcommand{\eeq} {\end{equation}}

\newcommand{\cb} {{c\parallel B_0}}

\newcommand{\lsco} {{La$_{2-x}$Sr$_x$CuO$_4$}} 
\newcommand{\ybco} {{YBa$_{2}$Cu$_3$O$_7$}} 
\newcommand{\pcco} {{Pr$_{2-x}$Ce$_x$CuO$_4$}} 
 
\begin{document}


\title{Distribution of electrons and holes in cuprate superconductors as determined from $^{17}$O and $^{63}$Cu nuclear magnetic resonance}


\author{Michael Jurkutat}
\affiliation{University of Leipzig, Faculty of Physics and Earth Sciences,  Linnestr. 5, 04103 Leipzig, Germany}
\author{Damian Rybicki}
\affiliation{University of Leipzig, Faculty of Physics and Earth Sciences,  Linnestr. 5, 04103 Leipzig, Germany}
\affiliation{AGH University of Science and Technology, Faculty of Physics and Applied Computer Science,
Department of Solid State Physics, al. A. Mickiewicza 30, 30-059 Krakow, Poland}
\author{Oleg P. Sushkov}
\affiliation{University of New South Wales, Department of Physics, Sydney, Australia}
\author{Grant V. M. Williams}
\affiliation{School of Chemical and Physical Sciences, Victoria University of Wellington, PO Box 600, Wellington 6140, New Zealand}
\author{Andreas Erb}
\affiliation{Walther Meissner Institute for Low Temperature Research, Walther-Meissnerstr. 8, 85748 Garching, Germany}
\author{J\"urgen Haase}
\affiliation{University of Leipzig, Faculty of Physics and Earth Sciences,  Linnestr. 5, 04103 Leipzig, Germany}

\date{\today}

\begin{abstract}
The distribution of electrons and holes in the CuO$_2$ plane of the high-temperature superconducting cuprates is determined with nuclear magnetic resonance through the quadrupole splittings of $^{17}$O and $^{63}$Cu. Based on new data for single crystals of electron-doped Pr$_{2-x}$Ce$_x$CuO$_4$(x=0, 0.05, 0.10, 0.15) as well as Nd$_{2-x}$Ce$_x$CuO$_4$  (x=0, 0.13) the changes in hole contents $n_d$ of Cu 3d$(x^2-y^2)$ and $n_p$ of O 2$p_\sigma$ orbitals are determined and they account for the stoichiometrically doped charges, similar to hole-doped \lsco. It emerges that while $n_d+2n_p=1$ in all parent materials as expected, $n_d$ and $n_p$ vary substantially between different groups of materials. Doping holes increases predominantly $n_p$, but also $n_d$. To the contrary, doping electrons predominantly decreases $n_d$ and only slightly $n_p$. However, $n_p$ for the electron doped systems is higher than that in hole doped La$_{1.85}$Sr$_{0.15}$CuO$_4$. Cuprates with the highest maximum $T_{\rm c}$s appear to have a comparably low $n_d$ while, at the same time, $n_p$ is very high. The rather high oxygen hole content of the Pr$_2$CuO$_4$ and Nd$_2$CuO$_4$ with the low $n_d$ seems to make them ideal candidates for hole doping to obtain the highest $T_{\rm c}$.
\end{abstract}




\maketitle


The high-temperature superconducting cuprates (HTSCs) emerge from insulating antiferromagnetic parent materials, the essential unit of which is the CuO$_2$ plane. It is formed, in first approximation, by Cu$^{2+}$ ($3d^9$) and O$^{2-}$ ($2p^6$), with the half-filled Cu $3d({x}^2-y^2)$ orbital hybridized with four O $2p_\sigma$ orbitals, so that a near square planar arrangement of Cu and O follows. 
Charge neutrality is provided by layers between CuO$_2$ planes, the chemistry of which can be varied widely. By changing the average charge state of the layers, e.g., by doping them, electrons are added or removed from the CuO$_2$ plane. 
These additional electrons or holes destroy the Cu-based antiferromagnetism. 
Above a certain doping level superconductivity can emerge with a maximum critical transition temperature $T_{\rm c}$ at optimal average doping, and it decreases nearly parabolically as the doping departs from the optimal value. 
The highest $T_{\rm c}$'s vary widely among different materials, and it is not quite clear which material parameters affect this behavior. 
Therefore, the knowledge of the microscopic distribution of charges in the CuO$_2$ plane is of great interest for a better understanding of the doping process.

\begin{SCfigure}
   \centering
   \includegraphics[width=0.3\textwidth]{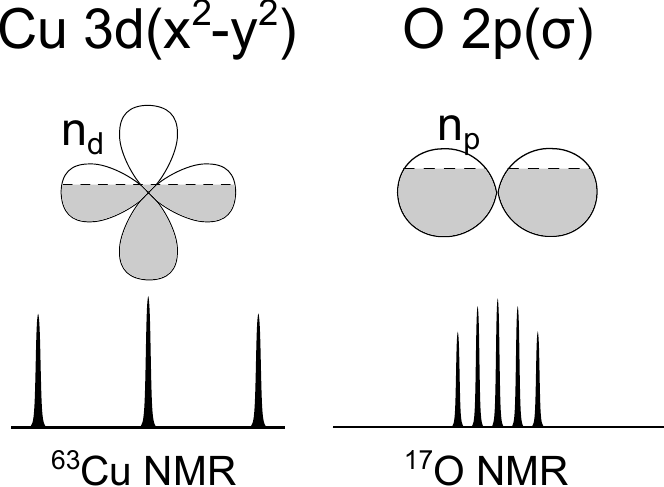}
   \caption{Holes in the orbitals cause a symmetric splitting of the $^{63}$Cu and {$^{17}$O} NMR into $2I=3$ and $2I=5$ lines, respectively.}
\label{fig:scheme}
\end{SCfigure}

A local probe like nuclear magnetic resonance (NMR) can yield vital information about the doping process since a change in the hole or electron content in the Cu $3d({x}^2-y^2)$ or O $2p_\sigma$ orbitals changes the electric field gradients (EFGs) experienced by the $^{63,65}$Cu ($I=3/2$) and $^{17}$O ($I=5/2$) nuclei through the electric quadrupole interaction that splits the nuclear resonances for Cu and O into $2I$ lines in a high magnetic field. Their frequencies are given by $\nu_n=\nu_0\pm n \nu_Q$: the central line ($n=0$) is in leading order unaffected by the quadrupole interaction that causes an angular dependent quadrupole splitting $\nu_Q$ for the $2I-1$ satellite transitions ($n=1$ for Cu and $n=1,2$ for O), cf. Fig.~\ref{fig:scheme}.

\begin{figure*}
\includegraphics[width=1\textwidth]{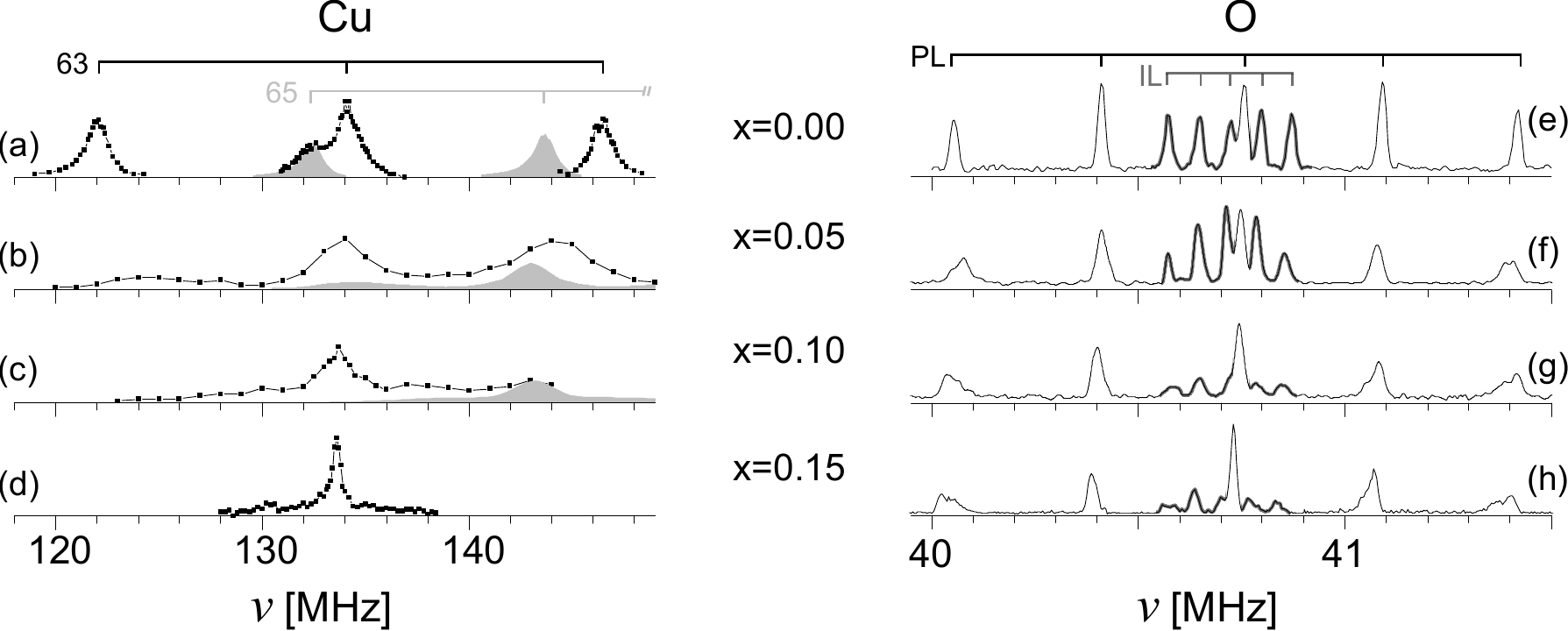}
\caption{(Color online) Spectra of Pr$_{2-x}$Ce$_x$CuO$_4$ for $\cb$ of Cu at 11.74~T (\textit{left} (a)-(d)) and O at 7.05~T (\textit{right} (e)-(h)) for different doping ($x$) ranging from undoped (\textit{top}) to optimally doped (\textit{bottom}). The spectra were measured at room temperature, except for (a) which was recorded at 400~K.
Inferred overlapping $^{65}$Cu spectra are indicated in gray in (a-c). 
The spectra of interlayer O are indicated by a thicker line in (e-h); the differences in spectral intensities for planar and inter-layer (IL) O are due to excitation conditions optimized for planar (PL) O (selective excitation of the central transition). Relative NMR signal intensity of inter-layer and planar O was confirmed to match relative abundance $S(IL)/S(PL) \approx 1$ with separate experiments at all doping levels.}
\label{fig:Spectra} 
\end{figure*}

It was reported early on from NMR experiments that the quadrupole splittings at the planar Cu and O nuclei increase approximately linearly with the doping of excess holes.\cite{Ohsugi1994, zheng_local_1995, Gippius1997} 
However, for doping with electrons very little is known since only a few reports on $^{63,65}$Cu NMR are available,\cite{abe_cu-nmr_1989, zheng_nmr_1989, yoshinari_antiferromagnetic_1990, kambe_1991, otha_1992, imai_superconductivity_1995, zheng_fermi-liquid_2003, mikhalev_temperature_2004, williams_carrier_2004, jurkutat_charge_2013} which do not even find a well-defined quadrupole splitting for doped compounds.
$^{17}$O NMR data for the electron doped materials are not available at all.

Here we report on new results of $^{63}$Cu and in particular the first $^{17}$O NMR of electron doped cuprates, Pr$_{2-x}$Ce$_x$CuO$_4$(x=0, 0.05, 0.10, 0.15) and Nd$_{2-x}$Ce$_x$CuO$_4$ (x=0, 0.13) single crystals. 
Based on these measurements and published data for the hole-doped materials we develop the analysis in Ref.~\cite{haase_planar_2004} further, such that we can quantitatively determine the hole contents of the Cu 3$d({x}^2-y^2)$ and O $2p_\sigma$ orbitals for all cuprates. This includes, in particular, the hole/electron distribution in the parent compounds, where we find that the inherent Cu$^{2+}$ hole is shared between Cu 3$d({x}^2-y^2)$ and O $2p_\sigma$ depending on the material chemistry, as suggested recently \cite{Rybicki2014}. Thus, we can relate the hole content of both orbitals quantitatively to the critical temperature.

The growth of the single crystals employed the Traveling Solvent Floating Zone technique \cite{lambacher_advances_2010}. The $^{17}$O exchange (48h in a $^{17}$O atmosphere of about 0.7 bar at 850$^\circ$C) was followed by the annealing of excess O \cite{williams_gap_2002}.
The NMR measurements were performed at magnetic fields of 7.05~T or 11.74~T, at room temperature and at 400~K for Cu NMR for the parent compound to be well above the magnetic ordering temperature. 
The NMR spectra were taken by either using a frequency stepped spin-echo (broad $^{63}$Cu lines) or by adding up Fourier transform spectra (narrow $^{17}$O lines). Population transfer double-resonance \cite{haase_new_1998} was used for relating the various transitions to one another.

First, we address the $^{63}$Cu NMR. 
Typical spectra as a function of electron doping are displayed in Fig.~\ref{fig:Spectra}, showing a well-resolved quadrupole splitting in the parent compound, which decreases and is smeared out upon doping. 
The splitting in the parent compound (top left) is about 12.2(1)~MHz with $B_0$ along the crystal c-axis. 
This is similar to what has been reported for the related compound Nd$_2$CuO$_4$, earlier \citep{kumagai_cu-nmr_1989, kohori_cu_1989, abe_cu-nmr_1989}. 
A linewidth of about 1.2(1) MHz for all transitions, as expected for dominating magnetic inhomogeneity that affects all transitions equally, is also similar to what has been reported previously for the paramagnetic phase (above $T_N$) \cite{Pozzi1999}. 
The quadrupole splitting decreases approximately linearly with doping, while the sattelites broaden rapidly already at low doping. The central transition width, in contrast, decreases. 
This means that the character of the inhomogeneity changes from magnetic to quadrupolar with doping since a distribution of EFGs does not affect the central line ($n=0$). 
The rather narrow central resonance with a broad background near optimal doping has been reported previously \citep{abe_cu-nmr_1989, zheng_nmr_1989, kambe_1991, imai_superconductivity_1995, zheng_fermi-liquid_2003, williams_carrier_2004} and was attributed to charge disorder. 
In a recent, extensive investigation [13] some of us confirmed that this background is indeed due to broadened satellites with an average splitting of $^{63}\nu_Q\approx 0$~MHz, but with an EFG that remains nearly symmetric, which is in accord with what we report here.

Next, we address the $^{17}$O NMR. 
Orientation dependent studies, relaxation measurements and double resonance experiments were employed (not shown) to disentangle the various (up to 15) transitions and assure the assignment of the lines to planar and inter-layer oxygen. 
Note, the apical oxygen site above or below planar Cu present in all hole doped systems is vacant in these electron doped, so-called T'-structures\cite{takagi_1989} where inter-layer O is located above and below planar O. 
\begin{figure}
\includegraphics[width=0.5\textwidth]{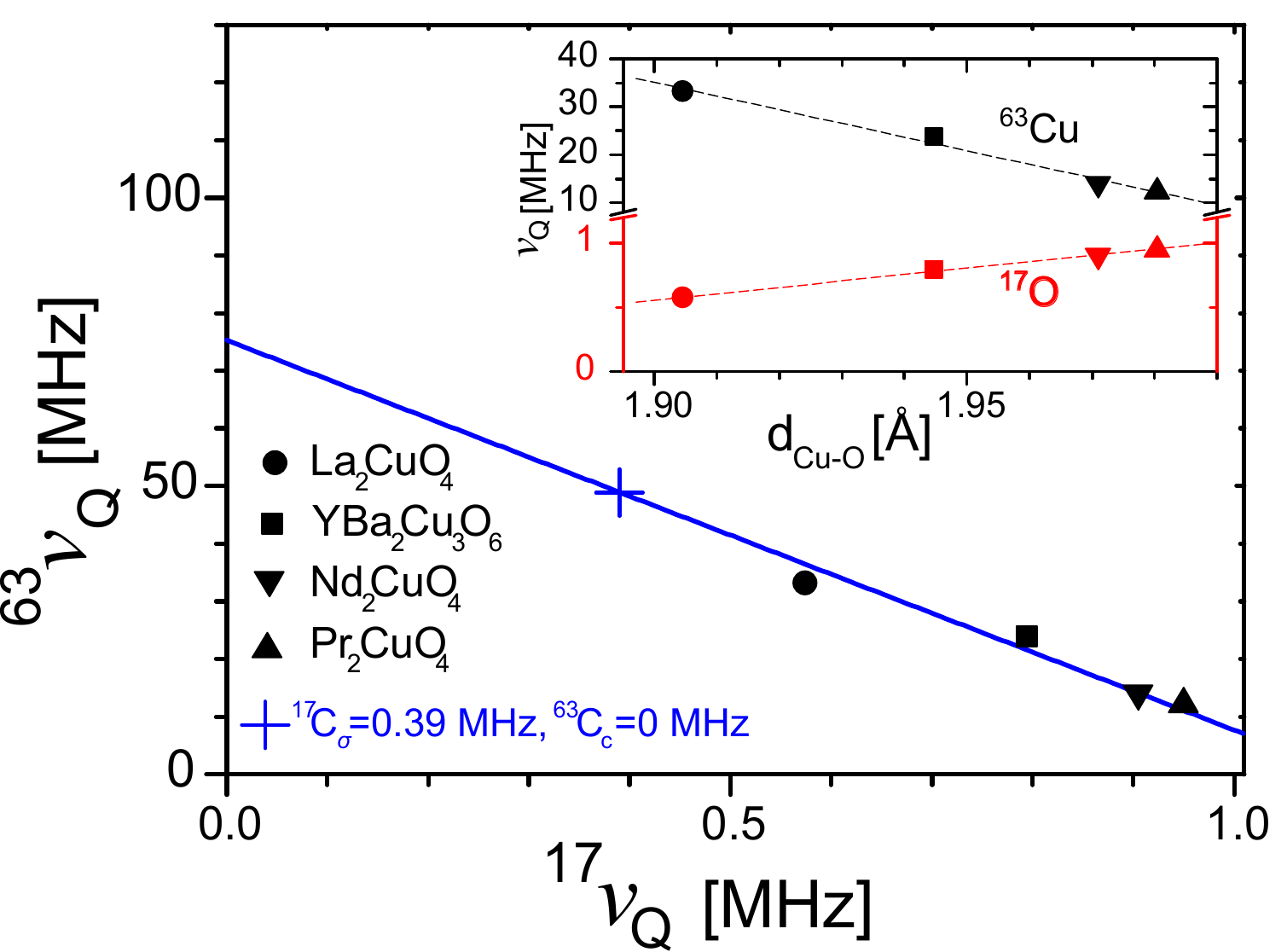}
\caption{(Color online) Planar Cu quadrupole splitting ($\cb$) vs. that of planar O ($\sigma \parallel B_0$) for undoped CuO$_2$ planes of HTSC parent compounds. Also shown is the expected dependence (
blue line) from a redistribution of the Cu $3d$ hole, for details see text. \textit{Inset}: $^{17}$O (full red symbols) and $^{63}$Cu (full black symbols) splittings from the main panel vs. planar Cu-O distance, dashed lines are guides to the eyes.}
\label{fig:undopedCuO2}\end{figure}
Typical spectra with the magnetic field $B_0$ perpendicular to the CuO$_2$ plane are shown in the right panel of Fig.~\ref{fig:Spectra}. 
The planar O has a rather large splitting in the parent material (larger than that of planar O in optimally \textit{hole} doped La$_{1.85}$Sr$_{0.15}$CuO${_4}$  and similar to optimally doped \ybco \cite{haase_planar_2004}), and it decreases only slightly with electron doping. The linewidths point to a doping-induced quadrupolar broadening, however, much weaker compared to planar Cu. 
The splitting of the inter-layer O is rather small as expected for a nearly full shell.

The planar Cu \cite{jurkutat_charge_2013} and O splittings show that the doped electrons enter predominantly the $3d({x}^2-y^2)$ orbital, as one might have expected. 
However, the rather large, doping-independent planar O splittings in Pr$_{2-x}$Ce$_x$CuO$_4$ (and Nd$_{2-x}$Ce$_x$CuO$_4$, spectra not shown) raise the question whether they reflect a large hole content or if they are due to other effects. 
It appears unlikely that the EFG is due to background charges that accidentally have a rather large influence only in these systems. 
In addition,  the fact that the planar O splittings are hardly affected by the strong Cu inhomogeneity and the proximity of an increasing number of doped ions also indicates that the background plays no role.

In order to shed light on this issue, we present in Fig.~\ref{fig:undopedCuO2} literature data for the quadrupole splittings of Cu and planar O for various \textit{parent} materials of hole and electron doped systems (where data were available). 
In the inset we plot the splittings vs.~the in-plane Cu-O distance. 
The Cu splitting decreases linearly with distance, while that for planar O increases linearly, as well. 
Hence, $^{63}\nu_{\rm Q}$ must be a linear function of $^{17}\nu_{\rm Q}$, and this is plotted in the main panel. 
While the dashed lines in the inset are just guides to the eye, the blue solid line in the main panel follows from a known model calculation \cite{haase_planar_2004} that we discuss now.

In this model, atomic spectroscopy data for the electric hyperfine interaction of isolated ions are used to estimate the doping dependent EFGs as being due the hole content of the Cu 3$d({x}^2-y^2)$ ($n_d$), and the O $2p_\sigma$ ($n_p$) orbitals. 
Basically without adjustable parameters this model reproduced the average number of doped holes $x=n_d+2n_p-1$ per CuO$_2$ of La$_{2-x}$Sr$_{x}$CuO${_4}$. 
It was found that mostly $n_p$ increases with increasing $x$.
In this model, the splittings, $^{63}\nu_{Q,c}$ (with the magnetic field along the crystal c-axis) and $^{17}\nu_{Q,\sigma}$ (with the magnetic field along the $2p_\sigma$ bond) are given by,
\begin{equation}
\begin{split}
&^{63}\nu_{Q,c}= 94.3 \mathrm{MHz} \cdot n_d - 14.2 \mathrm{MHz} \cdot \beta^2(8-4n_p) + {^{63}C_c} \\
&^{17}\nu_{Q,\sigma}=2.45 \mathrm{MHz} \cdot n_p +{^{17}C_\sigma}. \label{eq:HS}
\end{split}
\end{equation}

The first term on the r.h.s. of either equation relates to the hole content in bonding orbitals for Cu and O. 
The second term for $^{63}\nu_{Q,c}$ reflects the effect of Cu $4p$ occupation arising from overlap with O $2p_\sigma$ (of the 4 surrounding planar O), characterized by the constant $\beta^2 \approx 0.4$. Note that charges in O $3s$ and Cu $4s$ do not contribute to the EFG.
The material specific constants ${^{63}C_c}$ and ${^{17}C_\sigma}$ were introduced to account for the differences in the splittings for different parent compounds.

Recently, some of us applied this model to an electron doped system for the first time (Pr$_{1.85}$Ce$_{0.15}$CuO$_4$). It was found that almost all doped electrons enter the Cu  $3d(x^2-y^2)$ orbital, which results in a decrease of $ n_d$ by 0.13 at $x=0.15$.
This is the reason for the almost vanishing Cu quadrupole splitting discussed above \citep{jurkutat_charge_2013}. 

We now extend the model \cite{haase_planar_2004} to various \textit{parent} materials, i.e., we inquire about the meaning of the constants ${^{63}C_c}$ and ${^{17}C_\sigma}$. 
For the undoped parent material we must have $x=0$, which requires $n_d + 2 n_p=1$. This constraint together with
 \eqref{eq:HS} leads to the linear equation,
\begin{equation}
^{63}\nu_{Q,c} = -67.62 \cdot {^{17}\nu_{Q,\sigma}} + {^{63}\nu_{Q,c}^0},
\label{eq:slope}
\end{equation}
with the additive constant defined by,
\begin{equation}
^{63}\nu_{Q,c}^0 = 67.62 \cdot {^{17}C_\sigma} + {^{63}C_c} + 48.86 \mathrm{MHz}.
\label{eq:offset}
\end{equation}
The (blue) solid line in Fig.~\ref{fig:undopedCuO2} has the slope given by \eqref{eq:slope}, a strong indication that the model describing the various parent compounds by varying $n_d$ and $n_p$ with material-independent $^{63}C_c$ and $^{17}C_\sigma$ is appropriate. In order to determine the absolute hole contents we need to determine these constants that are not uniquely defined by \eqref{eq:offset} and $^{63}\nu_{Q,c}^0=75.3$~MHz, the intersect in Fig.~\ref{fig:undopedCuO2}. 

First, we discuss $^{17}C_\sigma$. Here, it is well-known that the anisotropy of the oxygen EFG is not zero, which demands that there is a contribution other than that from a $2p_\sigma$ hole. 
In a recent publication, it was shown by analyzing all available literature data that the anisotropy of the oxygen EFG is rather similar for all cuprates (with slight differences for two different groups) \cite{Rybicki2014}. This includes the electron doped systems investigated here.
This anisotropy (0.24~MHz) gives a lower bound for $^{17}C_\sigma$. 
On the other hand, it was also shown that the maximum critical temperature ($T_{\rm c,max}$) of different hole-doped cuprates is a linear function of the oxygen splitting and that $T_{\rm c,max}=0$ follows for a splitting of 0.5~MHz \cite{Rybicki2014}, which must represent an upper bound, i.e., $0.5~\mathrm{MHz}>{^{17}C_\sigma} >0.24$~MHz. 
Next, we note that one expects ${^{63}}C_c$ to be negligibly small \citep{stoll_electric_2002}, and since ${^{63}}C_c = 0$ demands $^{17}C_\sigma=0.39~\mathrm{MHz}$, a value that is in the range mentioned above, we use this pair of constants (${^{63}C_c}$=0~MHz and ${^{17}C_\sigma}=0.39\mathrm{MHz}$, blue cross in Fig.\,\ref{fig:undopedCuO2}) for simplicity. We indicate where appropriate what a different choice will mean for the absolute hole content.
\begin{figure}
\includegraphics[width=0.42\textwidth]{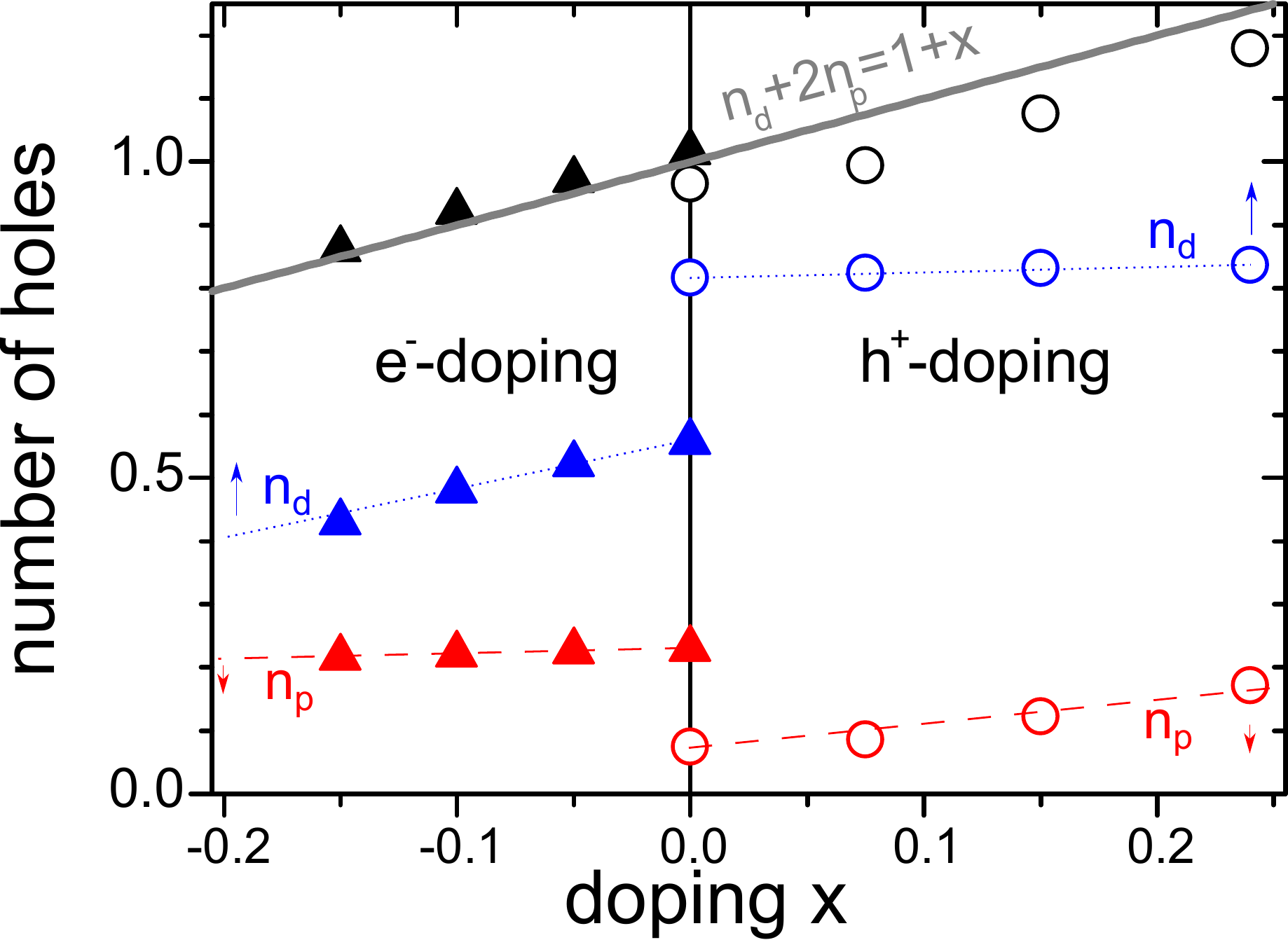}
\caption{\label{fig:4}(Color online) Doping dependence of the calculated number of holes  ($n_d$, blue symbols, indicated by dotted lines; $n_p$, red symbols, indicated by dashed lines) for hole-doped La$_{2-x}$Sr$_x$CuO$_4$ ($\medcircle$), and electron-doped Pr$_{2-x}$Ce$_x$CuO$_4$ ($\filledmedtriangleup$). Also shown the total experimental hole density (black symbols) along with that expected stoichiometry (solid grey line). The arrows indicate the shift of all data points for a different choice of constants (see text).}
\end{figure}
Thus, we can use \eqref{eq:HS} to determine $n_d$ and $n_p$, as well as the planar doping $n_d+2n_p-1$ for the doped systems.

In Fig.~\ref{fig:4} we show the extracted doping-dependent hole densities for \pcco\; and \lsco\; where the doping ($x$) is known precisely from the stoichiometry. (The solid arrows indicate the shifts of the absolute hole contents for a somewhat larger ${^{17}C_\sigma} = 0.49$~MHz.) Note that the solid line representing the overall planar hole density has no adjustable parameters, and its good agreement with the data further supports the validity the model.
\begin{figure}
\includegraphics[width=0.42\textwidth]{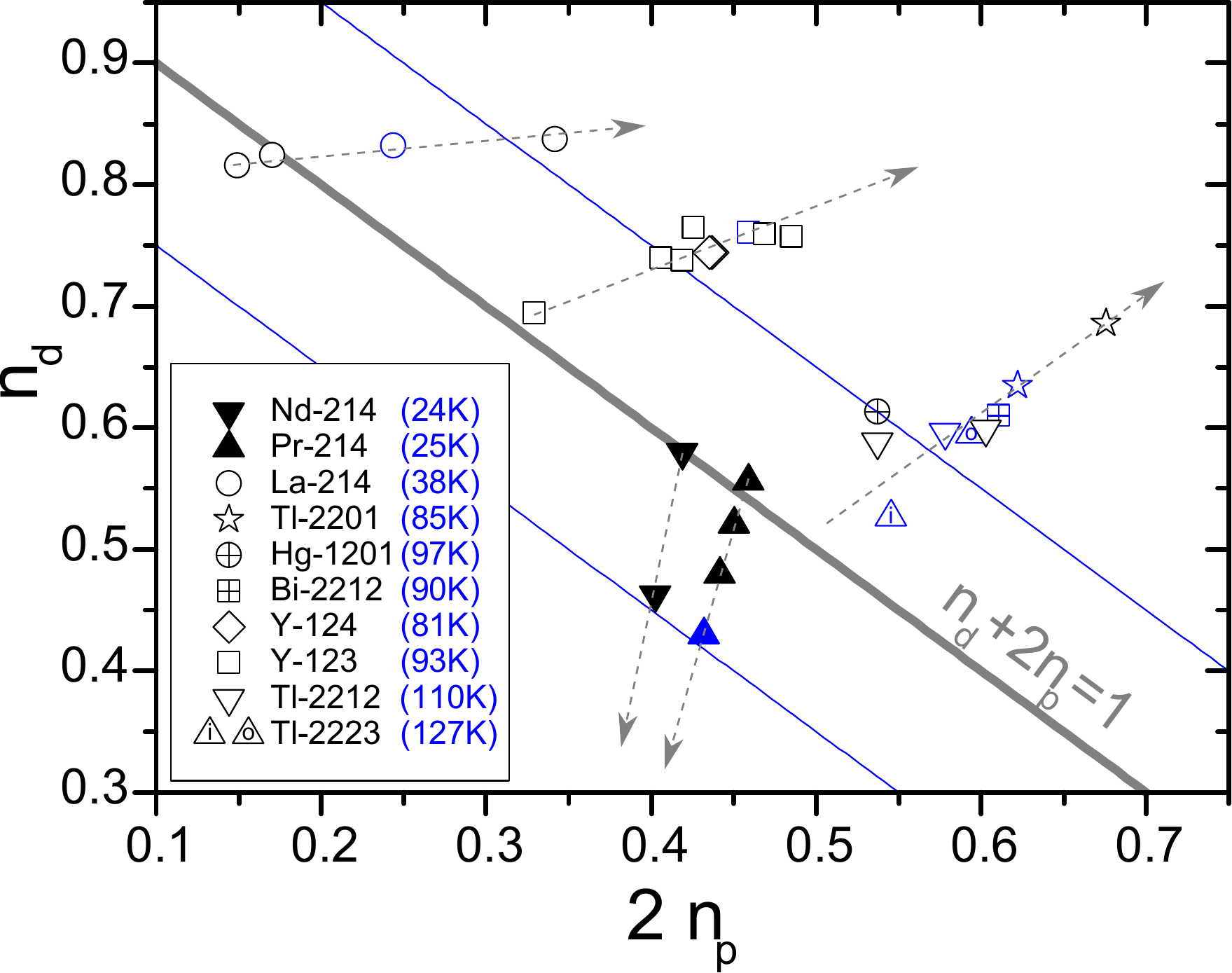}
\caption{\label{fig:5} (Color online) Planar Cu vs. O hole densities as extracted from quadrupole splittings for various HTSCs. In \textit{blue} we mark the maximum critical temperature ($T_{\rm c,max}$), data points and lines corresponding to optimal doping. The \textit{grey} line corresponds to the undoped case with 1 hole per CuO$_2$. Dashed arrows are guides to the eye indicating changes due to an increase in doping. Data for hole-doped compounds are from \cite{Rybicki2014}.}
\end{figure}

We conclude that at least for the systems shown, it is a very good approximation to determine the number of Cu and O holes just from the apparent quadrupole splittings. Contributions to the EFGs from the lattice must play a minor role, as suggested earlier \cite{stoll_electric_2002}. Thus, in the parent materials the Cu$^{2+}$ hole is shared between the 3$d({x}^2-y^2)$ and 2$p_\sigma$ orbitals, $n_d+2n_p=1$, with $n_d$ and $n_p$ depending on the material. Doping $x$ changes $n_d$ and $n_p$ further, such that $n_d+2n_p=1+x$. In particular, we conclude that the planar O in the electron doped materials has a rather high hole content that exceeds that in La$_{1.85}$Sr$_{0.15}$CuO${_4}$.

Qualitatively, the higher oxygen hole content in the parent compounds
for electron doped cuprates compared to that for hole doped
cuprates was predicted a long time ago~\cite{Ohta91,Ohta92}.
According to these works the hybridization between the Cu 3d electron and O 2p electron is stronger
in electron cuprates because of the smaller charge transfer gap.
The charge transfer gaps are different in these two families because of
different Madelung potentials due to different configurations of apical
oxygens.
In spite of the same qualitative trend, the value of the effect reported
in the present work is 4-5 times larger than the prediction~\cite{Ohta92}.


Clearly, with such a quantitative model at hand, one is tempted to interpret all planar Cu and O splittings in terms of hole and electron content of the 3$d({x}^2-y^2)$ and 2$p_\sigma$ orbitals, even if data for the corresponding parent materials are not available.  
We therefore have converted literature data gathered in a recent publication \cite{Rybicki2014} according to the model described above in terms of $n_d$ and $n_p$. 
The results are shown in Fig.~\ref{fig:5} where we plot the Cu hole content vs. twice the oxygen hole content. 
The data points for hole-doped (open symbols) and electron-doped (full symbols) are widely spread and show quite different distributions of the doped charges, as evidenced by the dashed grey arrows indicating the changes with doping, for the different families. 
Nonetheless, we find for all families a good agreement for the parent compounds with zero doping (thick solid gray line, $n_d+2n_p=1$). 
Also, data points corresponding to optimal doping (blue symbols) for the different families are near 15\% hole or electron-doping (thin solid blue lines, $n_d+2n_p=1+0.15$ and 1-0.15). 
In particular for those HTSCs where no parent data are available and the doping ($x$) is not readily extracted from the stoichiometry, this accordance, again, supports the validity of our approach.
 
The materials with maximum $T_{\rm c}$ are found for low $n_d$  and large $n_p$, and in which hole doping increases $n_p$ even further. Electron doping decreases predominantly $n_d$ and leaves a high $n_p$ unchanged. It appears that doping these T'-structure systems with holes might promise a high $T_{\rm c}$, as well. This trend that has the materials with higher $T_{\rm c}$ further to the lower right in Fig.~\ref{fig:5} correlates with band structure calculations and an increase of the apical oxygen distance from Cu, although a deeper understanding is lacking, \cite{pavarini_2001} (note that Cu 4s charges would not contribute to the EFG).

To conclude, we have presented new $^{63}$Cu and the first $^{17}$O NMR data on single crystals of electron doped cuprates. An analysis of these new data in conjunction with available literature data confirms that the experimentally determined quadrupole splittings measure the hole content $n_d$ and $n_p$ of the 3$d({x}^2-y^2)$ and 2$p_\sigma$ bonding orbitals, respectively. In the parent materials, where $n_d+2n_p=1$, we find that $n_d$ and $n_p$ vary for different systems and that those systems favor a high $T_{\rm c,max}$ that have a rather low $n_d$, but large $n_p$. Doping additional holes increases predominantly $n_p$, but also $n_d$, while doping with electrons predominantly decreases $n_d$ and only slightly $n_p$. For the two electron doped systems investigated, the hole content $n_p$ of planar oxygen remains suprisingly large. This suggests that a hole doping might lead to rather high $T_{\rm c,max}$, as well.

\begin{acknowledgments}
 
\textbf{Acknowledgement}
We are thankful to T. Tohyama and G. Khaliullin for important discussions
and M. Fujita for a useful communication. We also acknowledge financial support by Leipzig University, the DFG within the Graduate School Build-MoNa, the European Social Fund (ESF) and the Free State of Saxony.
 
\end{acknowledgments}

\bibliography{DR-PRL-doping}{}
\bibliographystyle{apsrev4-1}

\end{document}